\documentclass[amsmath,amssymb,reprint,superscriptaddress,nopacs,aps, prapplied,footnotebib]{revtex4-1}
\usepackage{graphicx,xcolor,float,hhline,braket}
\pagecolor[rgb]{1,1,1}
\def\darkmodecolor{\color{black}}

\begin{document}

%%% TITLE et al. %%%%
\title{\darkmodecolor Polarization agnostic continuous variable quantum key distribution}
\author{Brian P. Williams*}
\affiliation{Quantum Information Science Section, Oak Ridge National Laboratory, Oak Ridge, Tennessee USA 37831}
\email{williamsbp@ornl.gov}
\author{Nicholas A. Peters}
\affiliation{Quantum Information Science Section, Oak Ridge National Laboratory, Oak Ridge, Tennessee USA 37831}
\email{petersna@ornl.gov}

\begin{abstract}
\darkmodecolor
We introduce a polarization agnostic method for Gaussian-modulated coherent state (GCMS) continuous-variable quantum key distribution (CVQKD). Due to the random and continuous nature of the GCMS protocol, Alice, the transmitter, can encode two distinct quadratures in each of two orthogonal polarization modes, such that Bob, the receiver, measures valid GCMS quadratures in a single polarization mode even when polarization changes occur during transmission. This method does not require polarization correction in the optical domain, does not require monitoring both polarization modes, reduces loss by eliminating optical components, and avoids the noise injected by polarization correction algorithms. 
\end{abstract}

\maketitle
\darkmodecolor

Continuous variable quantum key distribution (CVQKD) based on optical coherent detection is a promising alternative to discrete photon methods due to its use of common telecom components, room temperature operation, ability to coexist with telecom traffic \cite{diamanti2015distributing,laudenbach2018continuous,williams2024field}, and advancing implementations with photonic integrated circuits \cite{bian2024continuous,hajomer2024continuous,pietri2024experimental}. 
The well-established Gaussian-modulated coherent state (GMCS) protocol \cite{grosshans2003quantum} has the most robust proven security and has been demonstrated over 200~km of fiber \cite{zhang2020long}. One technical challenge in CVQKD is to correct the polarization at the receiver such that the signal is aligned with Bob's local oscillator (LO), his reference laser used for homodyne measurement. When the polarization states of the signal and LO are not aligned Bob's signal-to-noise ratio is reduced. This reduction has a critical impact on the secret-key rate and achievable transmission distance.

Polarization correction and management strategies in CVQKD consist of passive and active methods. Polarization diverse receivers can be used to passively monitor both polarization modes \cite{pereira2023polarization,wang2019polarization,chu2020polarization,tan2024polarization,roumestan2024shaped} at the cost of additional detector resources. Active methods utilize a polarization controller and ancillary signal feedback to align the signal and LO \cite{williams2024field,liu2020continuous,guo2019polarization,wang2020high}. The reduction of detector resources achieved using active methods requires the addition of more complex optical modulation and real-time data processing. 

\begin{figure}[t]
\includegraphics[width=0.5\linewidth]{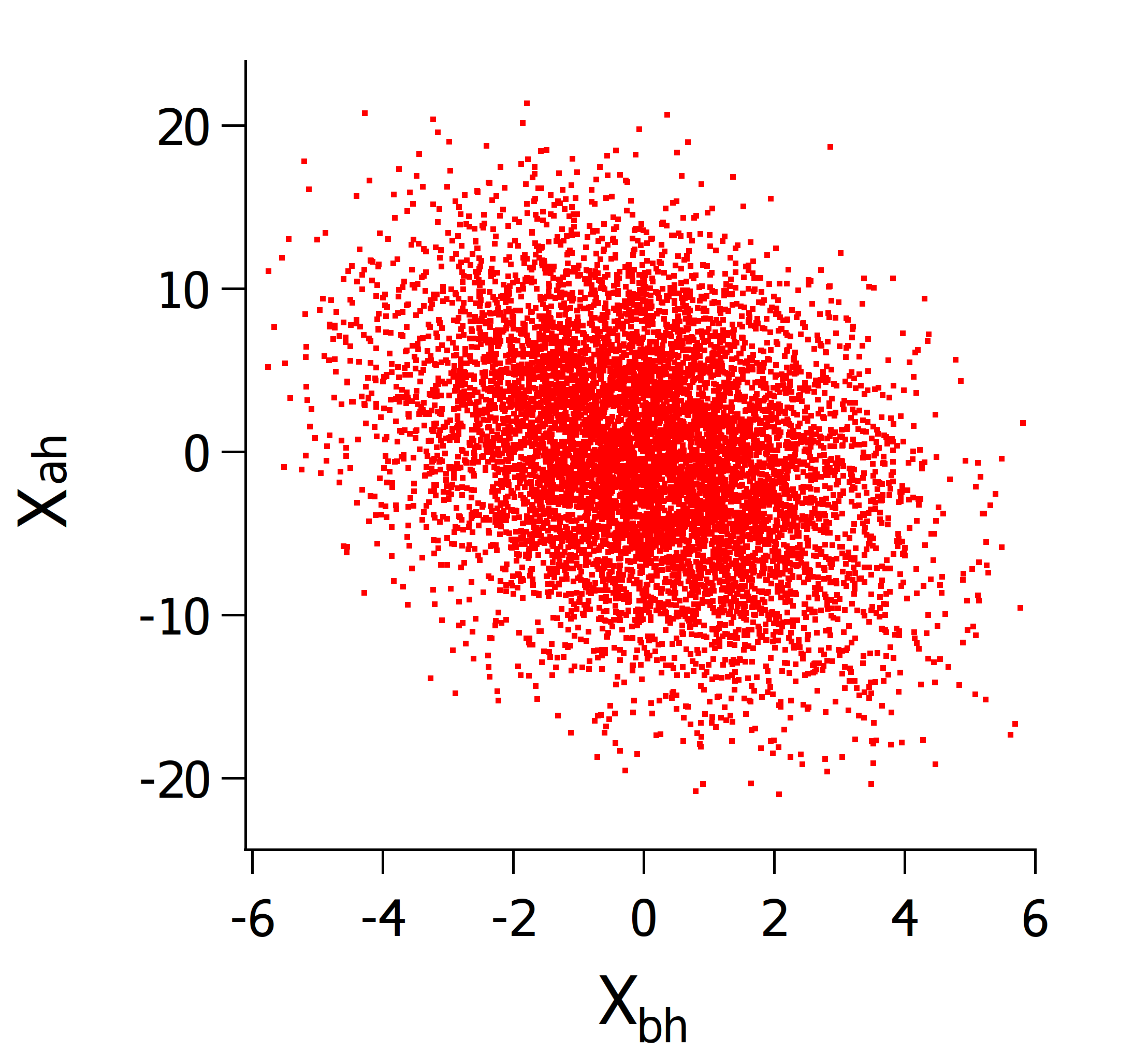}\includegraphics[width=0.5\linewidth]{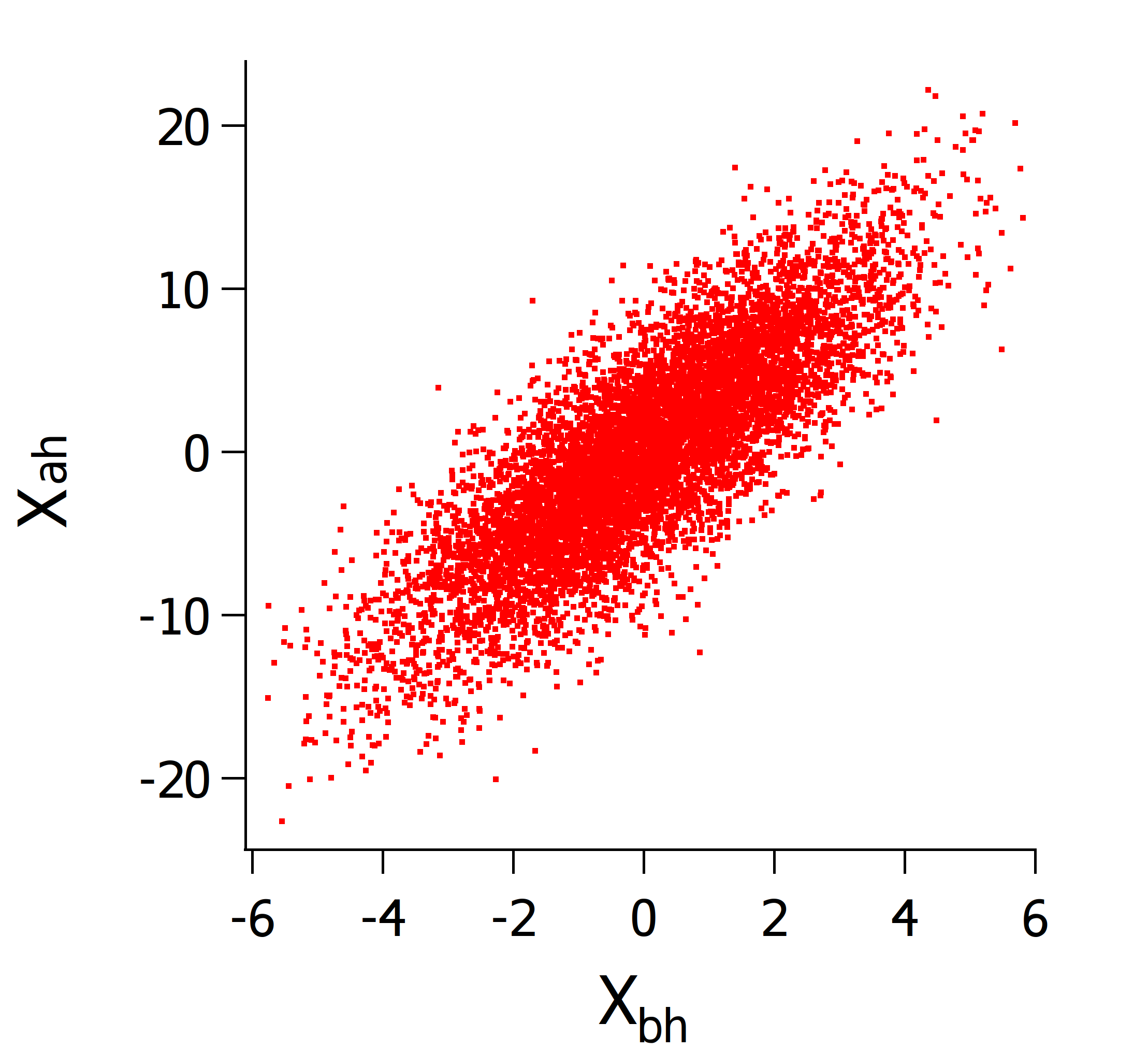}
\caption{Bob's $X_{bh}$ versus Alice's sent signal $X_{ah}$. Left) without corrections to Alice's encoding and Right) with corrections. Alice sends a GMCS signal with average photon number $\left\langle n \right\rangle = 5$, $V_A=10$. The polarization parameters recovered from this data are $\phi = 2.28$, $\delta=-0.05$, $\theta=0.84$.}
\label{corrected}
\end{figure}

In this article we present the first, to our knowledge, passive approach for GMCS CVQKD polarization management that does not require monitoring both polarizations at the receiver. Briefly, our Alice encodes independent GMCS signals in each of two orthogonal polarizations. These encoded values undergo a polarization transformation in the optical fiber but remain valid GMCS signals. Given that Alice can identify the polarization transformation post-transmission, Alice can digitally transform her stored encoding values such that correlations with Bob are recovered as seen in Fig. \ref{corrected}. Our method only requires monitoring of one polarization, eliminates the cost and loss associated with polarization control components, and avoids noise injected by polarization correction algorithms. 

\section{GMCS CV-QKD}

The fundamental make-up of a CV-QKD system is a coherent communications system \cite{yamamoto1981coherent,kikuchi2015fundamentals}. A weak signal, about one photon, is transmitted from Alice to Bob. Bob interferes this weak signal with a bright reference laser, the LO. This interference results in an amplification of the signal and allows a phase measurement relative to Bob's LO. Bob's measurement is often carried out using dual-quadrature homodyne detection which outputs two measurements, the quadratures $X$ and $P$ \footnote{Misleadingly this has been referred to as a heterodyne detector in the CVQKD community. In the communications industry heterodyne detection refers to using an LO of a different frequency than the signal  \cite{brunner2017low}}. The values $X$ and $P$ form the raw-key between Alice and Bob which feeds the reconciliation process to produce the final secret key.

A typical GMCS CVQKD protocol is carried out in the following steps:
\begin{enumerate}
\item Alice picks random quadrature values $X_a$ and $P_a$ from a common Gaussian distribution with variance $V_A$ in shot-noise units (SNU).
\item Alice encodes $X_a$ and $P_a$, in 
 $\sqrt{SNU}$, onto a light pulse which is equivalent to picking an average photon number and phase.
\item Alice transmits the encoded light pulse to Bob in a single-polarization through single-mode optical fiber in which polarization transformations occur.
\item Bob receives Alice's light signal. Bob corrects the polarization using active devices before detection. Alternatively, Bob can use a polarization diverse receiver to make quadrature measurements.
\item Steps 1-5 are repeated to complete a \emph{packet}. In our case 7.8k pulses are sent with 1 $\mu$s between pulses, 1 Mhz.
\item Bob sends Alice a fraction of his results which are utilized to estimate the transmission $T$ for the optical link and the excess noise $\xi$ both of which impact the final secret-key rate.
\item Alice and Bob generate a secret-key through a reconciliation protocol.
\end{enumerate}
%The coherent detection amplification process involving the LO results in an added noise in quadrature measurement. This noise is referred to as shot-noise. $X$ and $P$ values are normalized to the shot-noise, given in shot-noise units (SNU). If Bob does not correct the polarization of Alice's signal he would lose the polarization component that is not aligned to the LO polarization. Bob could utilize two LO's and two DHDs such that he can monitor in both polarizations to try and mitigate this loss. However, reconstruction of the signal from two measurement processes would include noise from two LOs. One-way to address this issue is to correct the polarization (align it with the LO) before measurement. It is also possible, that the LO could be aligned to the signal instead, to avoid signal loss.

The performance metrics for GMCS involve the variance of Bob's measurements and Alice's encodings for each quadrature. These variances are normalized to the measured quadrature shot-noise and given in SNU. The shot-noise is the variance due to the presence of the LO. It is typically measured, in part,  by blocking light entering the signal port of the detector. Bob's measured variance, in our experiment calculated for a single received packet, is given as
\begin{equation}V_B=\frac{\eta T}{2}\left(V_A+\xi\right)+1+\epsilon\label{Vb}\end{equation}
where $V_A$ is Alice's single quadrature modulation variance, $\xi$ is the excess noise assumed to be contributed by Eve, $\eta$ is the efficiency of Bob's detector, $T$ is the optical link transmission, the $\frac{1}{2}$ fraction results from the dual-quadrature homodyne detection, the ``1" term represents the shot-noise, and $\epsilon$ is the detector noise. Again, the variance quantities are all in SNU.

\section{Polarization Agnostic Encoding for GMCS CVQKD}

For the GMCS protocol, Alice independently samples values to encode quadratures $X,P$ from a common Gaussian distribution 
\begin{equation}
f\left(q\right)=\frac{1}{\sqrt{2\pi V_A}}e^{-\frac{q^2}{2 V_A}}\label{Gaussian}
\end{equation}
where $q\in\{X,P\}$.

Given sampled $X,P$ values, Alice's prepared signal can be written as the coherent state \cite{loudon2000quantum}
\begin{equation}
\left|\alpha\right\rangle=\hat{D}\left(\alpha\right)\left|0\right\rangle=e^{\alpha \hat{a}^\dagger -\alpha^{*}\hat{a}}\left|0\right\rangle
\end{equation}
where the creation and annihilation operators have the property $\left[\hat{a},\hat{a}^\dagger\right]=1$ and
\begin{equation}\alpha = \left|\alpha\right|\cos\gamma + \textrm{i} \left|\alpha\right|\sin\gamma= X+i P\textrm{.}\end{equation}

Using Jones polarization notation, Alice's state prepared in horizontal polarization can be written as
\begin{equation}\left| \alpha_H \textrm{,} 0 \right\rangle = \begin{bmatrix}
\alpha_h \\
0
\end{bmatrix}= \begin{bmatrix}
X_h+\textrm{i} P_h \\
0
\end{bmatrix}\textrm{.}\label{alpha_state}
\end{equation}
We note that this representation is a departure from some CVQKD literature where the components of the column vector are the quadratures $X,P$ and the polarization is assumed to be single-mode.

Due to the birefringent nature of standard single mode optical fiber, light transmitted through such fiber results in changes to the light polarization. It is shown in \cite{vanwiggeren1999transmission} that the many polarization rotations and phase changes accumulated in a singe-mode fiber transmission are equivalent to a single rotation
\begin{equation}R\left(\theta\right)=
\begin{bmatrix}
\cos\theta & -\sin\theta \\
\sin\theta & \cos\theta 
\end{bmatrix}\label{theta}
\end{equation}
and single phase change
\begin{equation}C\left(\delta\right)=
\begin{bmatrix}
e^{i\frac{\delta}{2}} & 0 \\
0 & e^{-i\frac{\delta}{2}} 
\end{bmatrix}
\end{equation}
given here as Jones matrices \footnote{We note that this does not fully account for temporal effects of phenomenon such as polarization mode dispersion (PMD).}.

In the case of CVQKD, a relative phase also exists between the LO and Alice's signal. This phase can be included as an additional global phase
\begin{equation}G\left(\phi\right)=
\begin{bmatrix}
e^{i\phi} & 0 \\
0 & e^{i\phi} 
\end{bmatrix}\textrm{.}
\end{equation}

Now, assume that Alice encodes independent quadratures in each polarization, all from a common Gaussian distribution. We can write a single instance of this encoding as
\begin{equation}\left| \alpha_H \textrm{,} \beta_V \right\rangle=\begin{bmatrix}
X_{ah}+\textrm{i} P_{ah} \\
X_{av}+\textrm{i} P_{av}
\end{bmatrix}\end{equation}
where $\alpha$ and $\beta$ describe the horizontal and vertical polarization coherent states, respectively.

The transformed state after propagating through a single-mode optical fiber is 
\begin{align}R\left(\theta\right)G\left(\phi\right)C\left(\delta\right)\left| \alpha_H \textrm{,} \beta_V \right\rangle=
\begin{bmatrix}
X_{bh} + \textrm{i} P_{bh} \\
X_{bv} + \textrm{i} P_{bv}
\end{bmatrix}\label{model}
\end{align}
where $X_{bh}$, $P_{bh}$, $X_{bv}$, and $P_{bv}$ depend on $X_{ah}$, $P_{ah}$, $X_{av}$, and $P_{av}$, for instance
\begin{multline}
X_{bh} = \left[X_{ah} \cos\left(\phi+\frac{\delta}{2}\right)-P_{ah} \sin\left(\phi+\frac{\delta}{2}\right)\right]\cos\theta\\
-\left[X_{av} \cos\left(\phi-\frac{\delta}{2}\right)-P_{av} \sin\left(\phi-\frac{\delta}{2}\right)\right] \sin\theta\textrm{.}\label{u0}
\end{multline}
Bob's quadratures $P_{bh}$, $X_{bv}$, and $P_{bv}$ have similar forms.
%\begin{align} \
%u_0 = &\left(X_0 \cos\frac{\delta}{2}-P_0 \sin\frac{\delta}{2}\right)\cos\theta-\nonumber\\
%&\left(X_1 \cos\frac{\delta}{2}+P_1 \sin\frac{\delta}{2}\right) \sin\theta\label{u0}\\
%v_0 = &\left(P_0 \cos\frac{\delta}{2}+X_0 \sin\frac{\delta}{2}\right)\cos\theta-\nonumber\\
%&\left(P_1 \cos\frac{\delta}{2}-X_1 \sin\frac{\delta}{2}\right) \sin\theta\label{v0}\\
%u_1 = &\left(X_1 \cos\frac{\delta}{2}+P_1 \sin\frac{\delta}{2}\right)\cos\theta+\nonumber\\
%&\left(X_0 \cos\frac{\delta}{2}-P_0 \sin\frac{\delta}{2}\right) \sin\theta\label{u1}\\
%v_1 = &\left(P_1 \cos\frac{\delta}{2}-X_1 \sin\frac{\delta}{2}\right)\cos\theta+\nonumber\\
%&\left(P_0 \cos\frac{\delta}{2}+X_0 \sin\frac{\delta}{2}\right) \sin\theta\label{v1} \textrm{.}
%\end{align}

Given that we rewrite $X_{ah},P_{ah},X_{av}$, and $P_{av}$ as functions of $X_{bh},P_{bh},X_{bv},$ and $P_{bv}$ using Eq. \ref{Gaussian} and \ref{model}, we can determine the $X_{bv}$ quadrature distribution by integrating over Bob's remaining quadratures $P_{bh},X_{bv},P_{bv}$ \footnote{The relevant Jacobian is equal to 1.}
\begin{align}F\left(X_{bh}\right)& =\int\int\int dP_{bh} dX_{bv} dP_{bv}\nonumber\\
&\times f\left(X_{ah}\right)\times f\left(P_{ah}\right)\times f\left(X_{av}\right)\times f\left(P_{av}\right)\textrm{.}
\end{align}

Alternative to solving Eq. \ref{model}, it may be verified that
\begin{equation} X_{ah}^2 + P_{ah}^2 +X_{av}^2 +P_{av}^2 = X_{bh}^2 + P_{bh}^2 +X_{bv}^2 +P_{bv}^2\label{equal}\end{equation}
such that
\begin{align}F&\left(X_{bh}\right) =\int\int\int dP_{bh} dX_{bv} dP_{bv}\nonumber\\
&\times f\left(X_{bh}\right)\times f\left(P_{bh}\right)\times f\left(X_{bv}\right)\times f\left(P_{bv}\right)\nonumber\\
&=f\left(X_{bh}\right)\label{independent}
\end{align}
resulting in the quadrature $X_{bh}$ having the same Gaussian distribution as Alice's original quadratures given in Eq. \ref{Gaussian}. Clearly, Bob's other quadratures $P_{bh}$, $X_{bv}$, and $P_{bv}$ also have this common Gaussian distribution. Above we have neglected loss and efficiency terms which will reduce the value of Bob's measured quadratures. This does not impact our result.

\begin{figure}[t]
\includegraphics[width=0.6\linewidth]{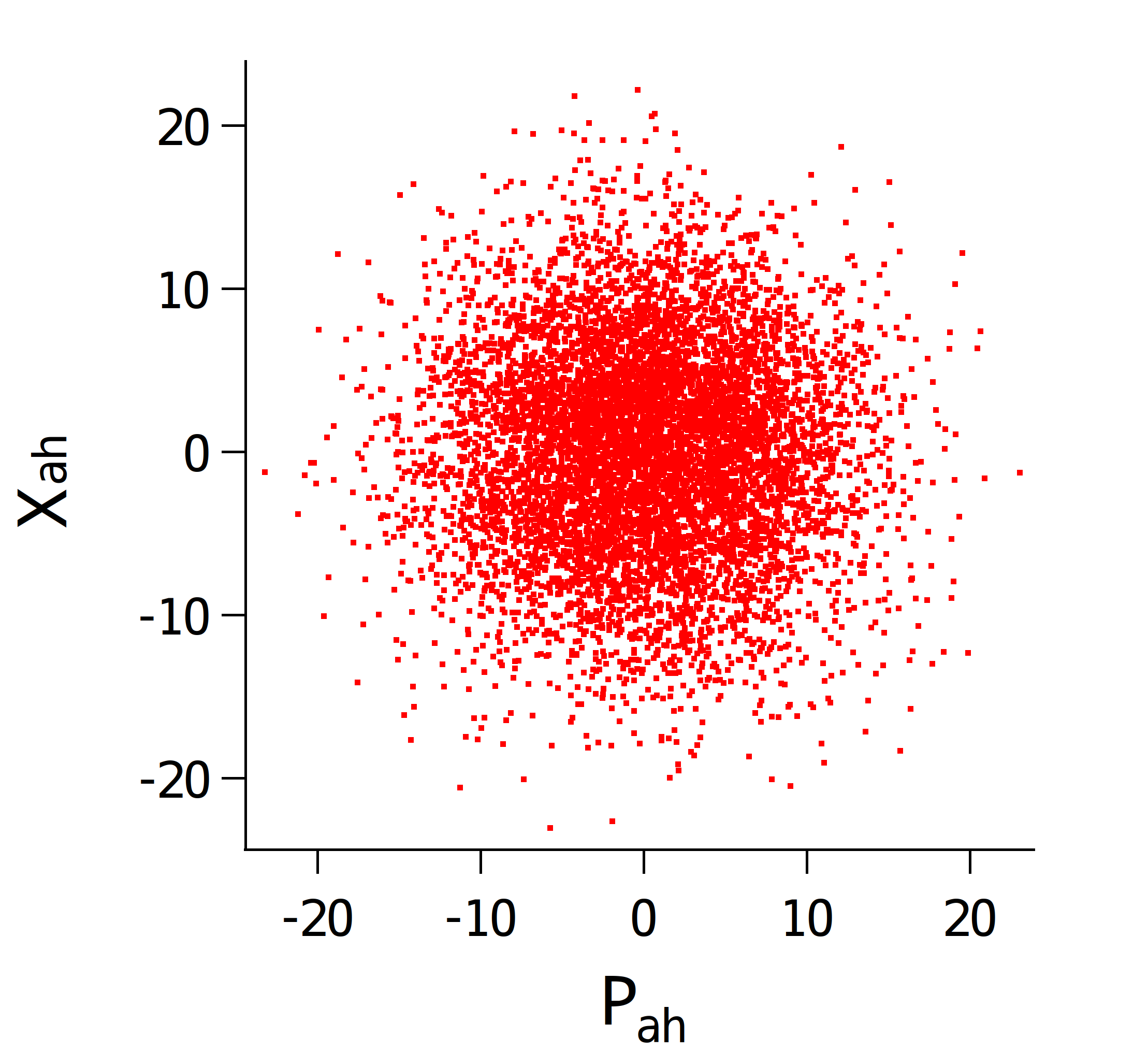}
\caption{Here we plot Alice's transformed values $X_{ah}$ and $P_{ah}$ to show that no correlations exist between transformed quadratures.}
\label{noCorrelation}
\end{figure}

Since polarization transformations result in Bob's quadratures being a combination of Alice's encoded quadratures, it seems plausible that there would be correlations in Bob's measurements. The calculation in Eq. \ref{independent} already supports quadrature independence, however for example, it can also be verified that
\begin{align} &\left\langle \left(X_{bh}-X_{bv}\right)^2\right\rangle \nonumber\\
&\quad=\int\int\int\int dX_{ah} dP_{ah} dX_{av}dP_{av}\nonumber\\
&\quad\quad\times f\left(X_{ah}\right)\times f\left(P_{ah}\right)\times f\left(X_{av}\right)\times f\left(P_{av}\right)\nonumber\\
&\quad\quad\times \left( X_{bh} - X_{bv} \right)^2 = 2 V_A
\end{align}
where $X_{bh},X_{bv}$ are functions of $X_{ah},P_{ah},X_{av},$ and $P_{av}$. This shows that no correlations exist between Bob's horizontal $X_{bh}$ and vertical $X_{bv}$ quadratures. The variance is just the sum of the individual Gaussian distributions. If there was a correlation, we would expect a reduction in this value. This same relation holds for any combination of the quadratures. Thus, there are no correlations between Alice's transformed quadratures or Bob's measured quadratures, which allows them to be treated as independent GMCS channels. As an example, we plot Alice's transformed values $X_{ah}$ versus $X_{av}$ in Fig. \ref{noCorrelation} from the experiment given in Fig. \ref{corrected}. As expected, no correlations appear.  

If Alice can determine with sufficient accuracy the polarization rotation and phase parameters $\theta$, $\phi$, and $\delta$ from Bob's shared data, she can digitally transform her recorded encoding values to match the optical polarization transformation with the transform given in Eq. \ref{model}. The final result will be as if Alice had originally encoded these quadrature values and no polarization changes occurred in the optical fiber link during transmission. It is not the case that Bob can correct his own dataset when using a single DQHD, since he does not have data from the unmonitored polarization.

Parameter estimation for the channel transmission $T$ and excess noise $\xi$ is a necessary step in the CVQKD protocol since they impact the achievable secret key-rate. As we will show, this same dataset can be used concurrently to determine the effective polarization parameters present in the signal received from Alice. 

Our modified polarization agnostic CVQKD protocol is:
\begin{enumerate}
\item Alice picks random quadrature values $X_{ah}$, $P_{ah}$, $X_{av}$, and $P_{av}$ from a common Gaussian distribution with variance $V_A$.
\item Alice encodes $X_{ah}$, $P_{ah}$ on the horizontal and $X_{av}$, $P_{av}$ on the vertical polarization components of her light pulse.
\item Alice transmits the encoded light pulse to Bob in a single-mode optical fiber in which polarization transformations occur.
\item Bob receives Alice's light signal and measures his own quadrature values $X_{bh}$, $P_{bh}$ in a single polarization.
\item Steps 1-5 are repeated to complete a \emph{packet}. In our case 7.8k pulses are sent with a 1 $\mu$s repetition rate, 1 Mhz.
\item Bob sends Alice a fraction of his results which are utilized to estimate the transmission $T$ for the optical link, the excess noise $\xi$, and the polarization parameters $\theta$, $\phi$, and $\delta$.
\item Alice rotates her stored quadratures digitally using estimates for $\theta$, $\phi$, and $\delta$.
\item Alice and Bob generate a secret-key through a reconciliation protocol.
\end{enumerate}

\section{Experiment}

\begin{figure}[h]
\includegraphics[width=0.7\linewidth]{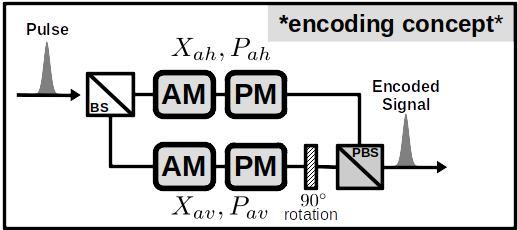}
\caption{Alice could encode each polarization in parallel utilizing four modulators. Here a beamsplitter (BS) splits her light pulse into horizontal and vertical encoding paths. An amplitude modulator (AM) and phase modulation (PM) in each path allows encoding $X_{ah}, P_{ah}, X_{av},$ and $P_{av}$ before using a half-wave plate with a 90$^\circ$ rotation and a polarizing beamsplitter (PBS) to recombine the encoded polarizations.}
\label{AliceExpt_two_paths}
\includegraphics[width=\linewidth]{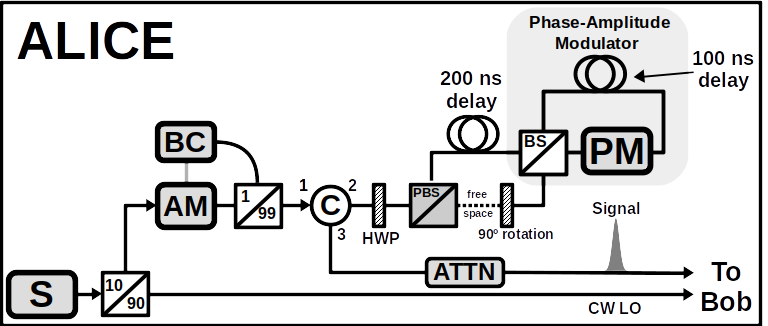}
\caption{Alice's PA-CV-QKD transmitter consists of a laser source (S) split into a signal and LO path using a 10:90 beamsplitter (BS). The signal pulse is generated using an amplitude modulator (AM) and sent to port 1 of the circulator (C). Exiting port 2 of the circulator the optical power is balanced equally into polarization modes using an adjustable half-wave plate. A polarizing beamsplitter (PBS) sends horizontal (H) and vertical (V) polarizations along different paths (of different lengths) to the phase-amplitude modulator where distinct quadrature values are encoded to $H$ and $V$. Encoded signals return to circulator port 2 and exit 3 to a final static attenuator (ATTN) before transmission to Bob. $10/90\equiv$ 10:90 BS; BC $\equiv$ bias controller; 1:99 $\equiv$ 1:99 BS; HWP $\equiv$ half-wave plate; PM $\equiv$ phase modulator; CW $\equiv$ continuous-wave; LO $\equiv$ local oscillator  }
\label{AliceExpt}
\end{figure}

One possible implementation of Alice's transmitter is seen in Fig. \ref{AliceExpt_two_paths} which utilizes parallel paths and four modulators to carry out encodings of all four quadratures. This transmitter is conceptually simple to understand and has the equivalent outcome as our method detailed below. 

To avoid the need to use four modulators, we implemented the transmitter seen in Fig. \ref{AliceExpt} which utilizes time-multiplexing and a bias-free phase-amplitude modulator \cite{williams2024field,dennis1996inherently,qi2018noise,Zhao:22} to encode each polarization's quadratures with a single LiNbO$_{3}$ waveguide phase modulator (EOSpace). Referencing Fig. \ref{AliceExpt}, a single frequency-stabilized continuous wave (CW) laser (OE Waves OE4028) generates 1542 nm light which is split using a 90:10 into the encoding and LO paths. The final CW LO power is 8.7 mW. In the encoding path, Alice utilizes a LiNbO$_{3}$ waveguide amplitude modulator (EOSpace) to generate 10 ns-wide pulses at a 1 MHz repetition rate.  After passing through a circulator (1$\rightarrow$2), a half-wave plate allows balancing of the light amplitude entering the horizontal and vertical polarization paths. The upper vertical path delays by 200 ns. The lower horizontal path has no delay but rotates the polarization from horizontal to vertical such that all encoded light travels the same optical path in the delay lines and phase-amplitude modulator. A phase modulator is placed asymmetrically in the Sagnac loop such that distinct phases can be applied to clockwise and counter-clockwise propagating pulses with a 100 ns delay between them \cite{williams2024field}. The encoded pulses ultimately exit the Sagnac loop along different delay paths than they entered removing the time delay between the different polarizations (overlapping them in time).  The encoded signal passes through the amplitude balancing HWP again which results in a change to the polarization state. This can be assumed to be part of the overall polarization transformation in the transmission link. After passing through the circulator (2$\rightarrow$3), a constant attenuation is applied to adjust the overall amplitude range before being sent to Bob.

\begin{figure}[tbh]
\includegraphics[width=\linewidth]{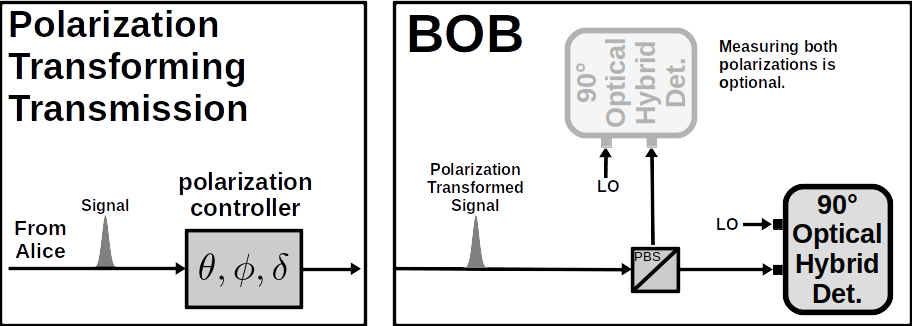}
\caption{Left) Between Alice and Bob we place a polarization controller to randomly change the polarization for each packet to simulate a real-world link. Right) Bob's CV-QKD receiver consists of a polarizing beamsplitter (PBS) and at least one DQHD (a 90$^\circ$ optical hybrid detector). One can monitor both polarization modes to double the key-rate at the cost of an additional DQHD.}
\label{BobExpt}
\end{figure}

Alice utilizes the transmitter in Fig. \ref{AliceExpt} to generate packets containing 7.8k 10-ns pulses with GMCS encodings in Alice's horizontal and vertical polarizations, 31.2k quadrature values per packet in total. A random polarization transformation is selected and applied to all the pulses in a single packet as seen at left in Fig. \ref{BobExpt}. We utilized a Luna PCD-M02 dynamic polarization controller to implement the polarization transformation. A unique random polarization transformation was applied to each packet. The packets contain 7.8k pulses at 1 $\mu s$ repetition rate, making the packet duration 7.8 ms. From the literature, a selection of reported polarization stability time durations range from $\mu s$ for coiled fiber in a high vibration environment \cite{Krummrich:04}, $ms$ in a 68 km of aerial fiber \cite{liu2019analysis}, to minutes in 55 km of underground optical fiber \cite{krummrich2005field}. Thus, the dataset used in parameter estimation should be adjusted accordingly. Clearly, parameter estimation will fail when the polarization changes approach the order of the signal repetition-rate. We note that polarization mode dispersion (PMD) is not typically relevant in current CVQKD systems due to their relatively slow operating speeds \cite{liu2019analysis}.

Bob's receiver consists of an LO and at least one dual-quadrature homodyne detector (DQHD), a 90$^\circ$ optical hybrid (90OH), as seen in Fig. \ref {BobExpt}. For this proof-of-principle experiment we are using Alice's laser as the LO, however a more secure method in a real deployment would be to utilize a true local oscillator \cite{qi2015generating,williams2024field}. At the cost of an additional DQHD, Bob has the option of measuring in both polarizations to double the key rate. In this case Bob's receiver would resemble a polarization diverse approach \cite{pereira2023polarization,wang2019polarization}, but for a given repetition rate we would have twice the key-rate. This doubling is due to having a unique GMCS CVQKD channel in each received polarization.

Bob measures all the pulses in a packet and extracts quadrature values $X_{bh},P_{bh}$ for each. Bob then sends Alice a portion of his measurements which Alice utilizes for parameter estimation. In this experiment, Bob shared 10$\%$ of his measured values. Utilizing a numerical least squares fitting algorithm \footnote{We used the C++ library Dlib, dlib.net.} and the model given by Eq. \ref{model} with terms such as those given in Eq. \ref{u0}, the parameter estimation process recovers the transmission T, excess noise $\xi$, and the polarization parameters $\theta$, $\phi$, and $\delta$.

\begin{figure}[t]
\includegraphics[width=0.49\linewidth]{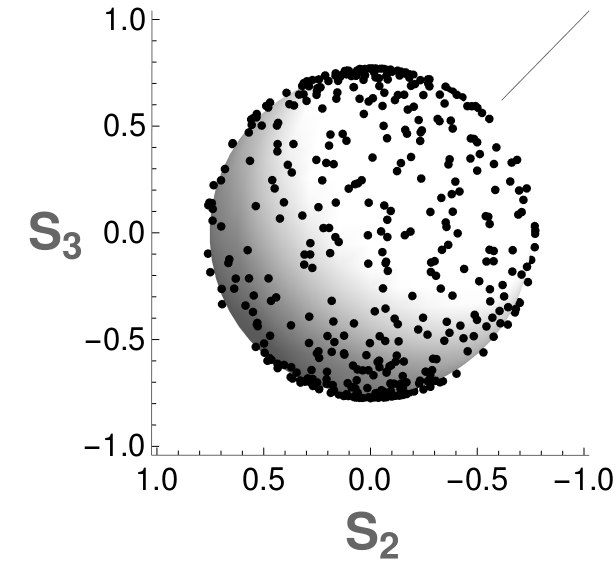}\;\;\includegraphics[width=0.49\linewidth]{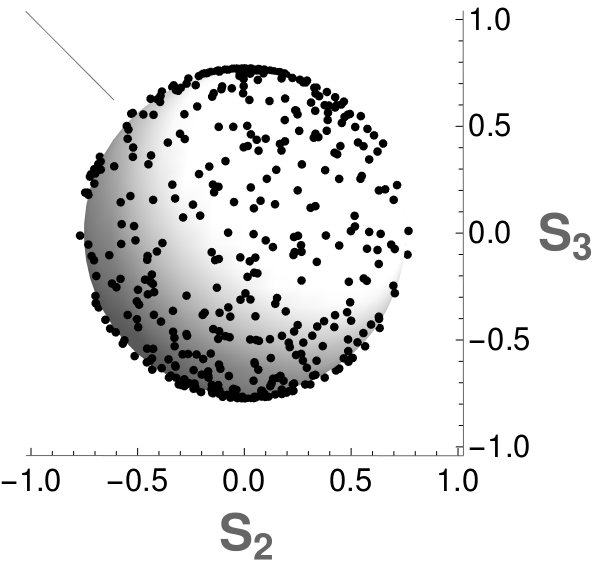}
\caption{Poincar\'e sphere representation of the Stokes parameters calculated using the polarization parameter estimates for $\delta$ and $\theta$ in our 1000 packet experiment.  The parameter $\phi$ is a global phase and is ignored for this visualization. The left and right plots give opposite views of the sphere. The Stokes parameter $S_1$ is into the page.}
\label{3dplot}
\end{figure}

To demonstrate our method we generated random polarization transformations (random $\theta,\phi,\delta$) for each of 1000 packets. Since our proof-of-principle table-top experiment is assumed to have $T\approx 1$ over a short (40m) optical fiber link, we used a relatively low modulation variance $V_A=1.16$, average photon number $\left\langle n \right\rangle =  V_A/2 = 0.58 $, to simulate real-world receiver side statistics. In Fig. \ref{3dplot} we give Poincar\'e spherical plots of the Stokes parameters given parameter estimates for $\delta$ and $\theta$. Our Stokes parameters \cite{collett2005field} are $S_1=-\cos\delta \sin 2\theta$, $S_1=\cos\delta \cos 2\theta$, and $S_3=\sin\delta$. As seen in Fig. \ref{3dplot} our random transformations extend over the full range of possible values. 

We use the $R^2$ value to characterize the correlations between Alice and Bob. In the case of our experiment this value is
\begin{equation}
R^2 = 1- \frac{1}{N\; V_B}\sum_{i=1}^N\left(X_{Bi}-\frac{\eta T}{2} X_{Ai}\right)^2
\end{equation}
where $V_B=1.32$, see Eq. \ref{Vb}, detector efficiency $\eta=0.5$, transmission $T=1$, excess noise $\xi=0.0328$, and detector noise $\epsilon=0.024$. It is straightforward to show that $R^2$ has maximum and minimum values
\begin{align}R^2_{\textrm{max}} &= \frac{\eta T V_A}{2 V_B}=0.219 \label{max}\\
R^2_{\textrm{min}} &= -\frac{\eta T V_A}{2 V_B}=-0.219\textrm{.}
\end{align}
%a maximum value of $R^2_{\textrm{max}} = 1 -\frac{1+\xi_d}{V_B} = 0.24$ and minimum value $R^2_{\textrm{min}} = 1 -\frac{2 \eta T V_A+\xi_d}{V_B}= -0.24$ given the experimental values for $V_A$, $\eta$, $T$, $\epsilon$, and detector noise $\epsilon$.

\begin{figure}[b]
\includegraphics[width=0.9\linewidth]{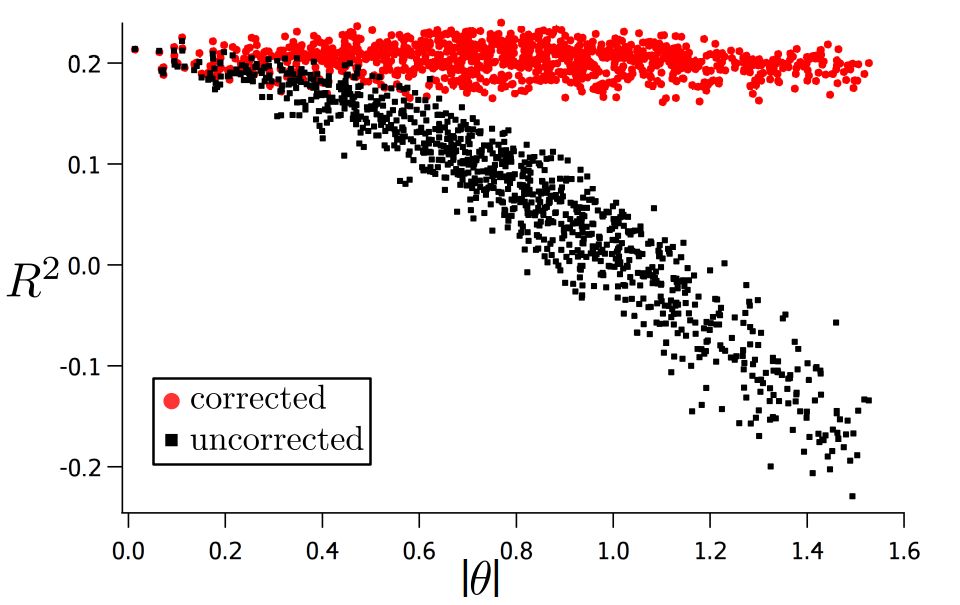}
\caption{$R^2$ correlation for $X_{ah}$ and $X_{bh}$ with and without rotating Alice's encodings with the estimated $\theta$ value. As expected, a high correlation is achieved when full corrections are made to Alice's encoding values. A decaying correlation is found when polarization rotations are ignored and not corrected.}
\label{R2}
\end{figure}

In Fig. \ref{R2} we plot data for the correlation between Alice's $X_{ah}$ and Bob's $X_{bh}$ quadratures versus polarization parameter $\theta$. Alice's encodings have been corrected for phase for all cases with the determined $\phi$ and $\delta$. The remaining parameter $\theta$ associated with the rotation given in Eq. \ref{theta} describes how much light rotates into the polarization orthogonal to a signal's original encoding polarization. When $\theta=0$ all signals stay in their original encoded polarization. For $\theta=\frac{\pi}{2}$ all light is rotated into the orthogonal polarization. In the case that Alice applies no $\theta$ correction to her stored encoding we find a correlation that decays with increasing $\theta$ reaching the predicted minimum at $\theta=\frac{\pi}{2}$. When Alice applies the correction for $\theta$ we measure a stable $R^2$ across all $\theta$ values with mean value $\langle R^2\rangle=0.214\pm0.005$, consistent with the expected value given in Eq \ref{max}. 

\begin{figure}[tbh]
\includegraphics[width=0.9\linewidth]{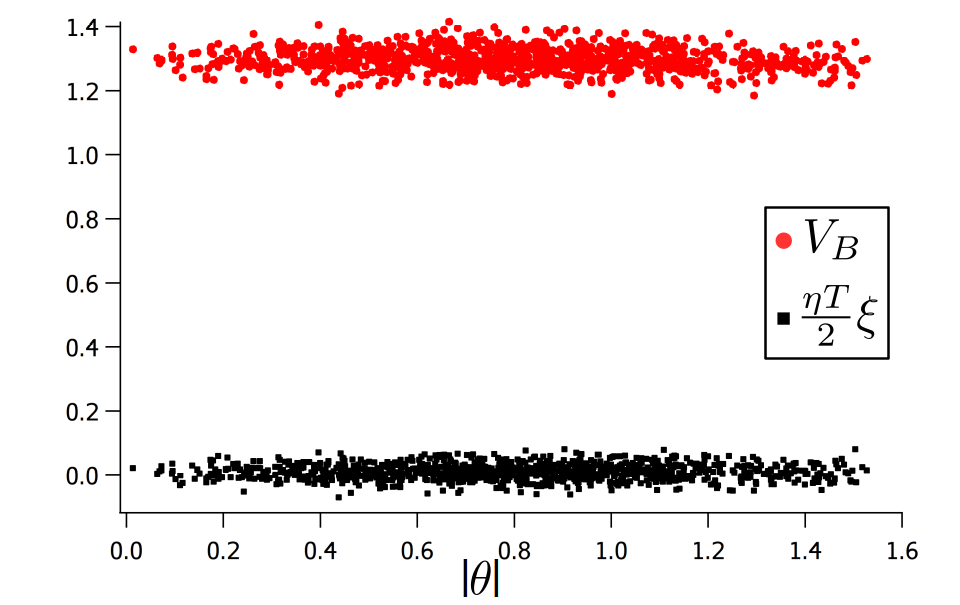}
\caption{Bob's measured variance $V_B$ and excess noise dependent quantity $\frac{\eta T}{2}\xi$ for quadrature $X_{bh}$. The $V_B$ data suggests that the received signal strength is maintained regardless of polarization changes. Likewise, we see no polarization dependence in the excess noise $\xi$.}
\label{VB}
\end{figure}

A key element of our method is that it maintains a constant average photon number $\left\langle n \right\rangle=V_A / 2$ from Alice even when polarization changes occur. This is important, since the key rate is maximized for a given transmission $T$ and excess noise $\xi$ only when a specific $V_A$ is utilized. Thus, the polarization changes impacting the signal will have no effect on the secret key-rate. As can be seen in Fig \ref{VB} Bob's measured variance $V_B$, given in Eq. \ref{Vb}, is independent of the polarization rotation parameter $\theta$. Similarly, we see no correlation in Bob's excess noise $\eta$ versus $\theta$. This demonstrates that corrections to Alice's stored encodings do not add polarization correction dependent noise which would manifest as excess noise. We note that the measured excess noise $\frac{\eta T}{2}\xi$ is plotted, key-rates are calculated using the noise we attribute to Eve which is the larger value $\xi$. The experimental excess noise from this data to be $\xi=0.0328\pm0.003$ which is similar to our previous work with the same devices \cite{williams2024field}.

\begin{figure}[t]
\includegraphics[width=0.8\linewidth]{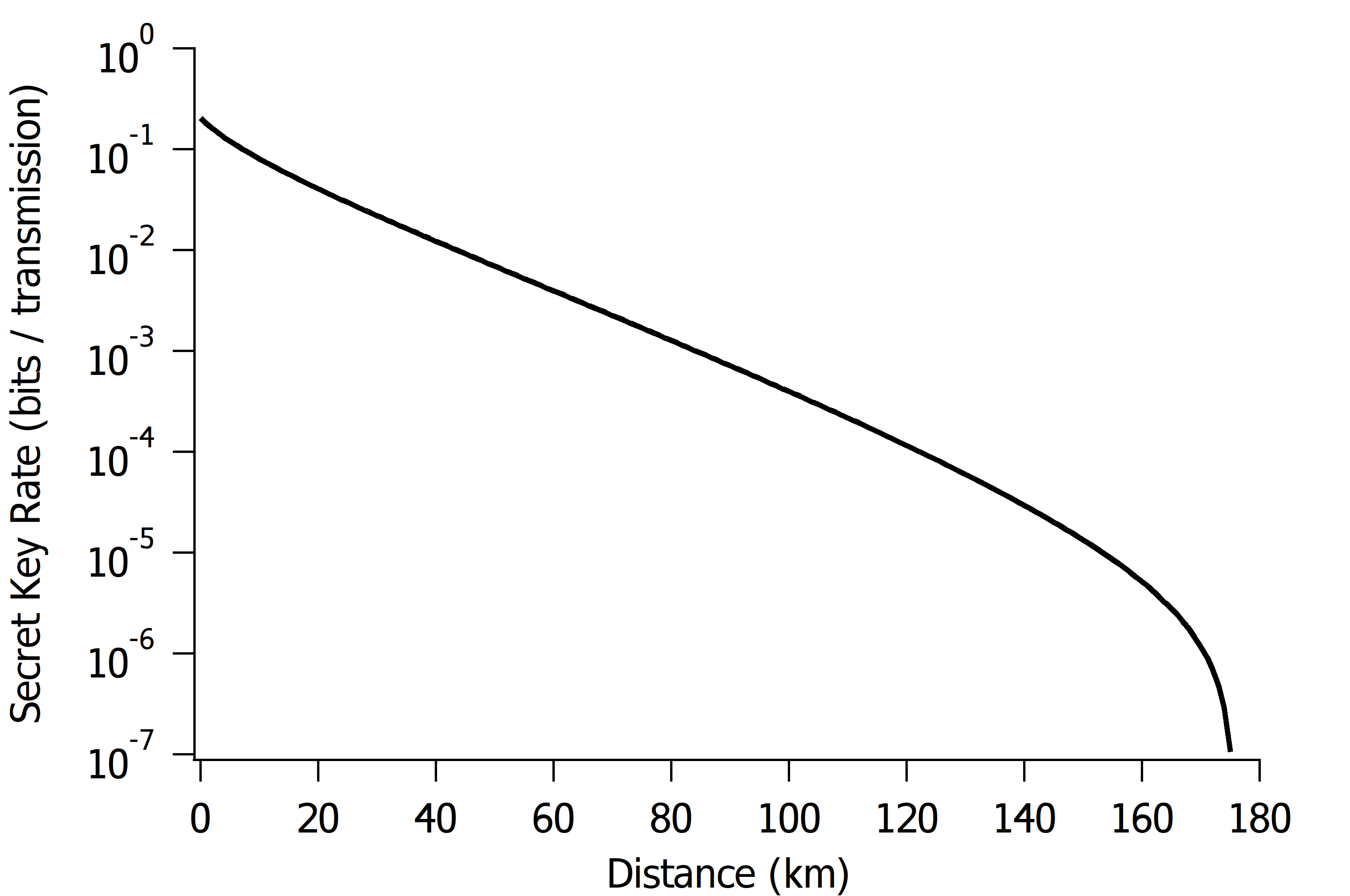}
\caption{Theoretical secret key-rate for the $X_{ah}$ quadrature given the experimental parameters and $V_A = 1$ with an infinite key-length and trusted detector.}
\label{keyRate}
\end{figure}

Given the measured excess noise $\xi=0.0328$, detector noise $\epsilon=0.024$, detector efficiency $\eta=0.5$, reconciliation efficiency $\beta=0.95$, and fixed modulation variance $V_A=1.16$, the secret-key rate prediction given an infinite key length and trusted detectors \cite{lodewyck2007quantum,fossier2009improvement} is given in Fig. \ref{keyRate}. We note that finite-size effects should be considered in a real deployment as it sets a scale for how many final keys may be generated in a given time period for a given configuration~\cite{leverrier2010finite}.

\section{Security implications}
While our method does introduce changes to Alice's parameter estimation process there are no changes that alter the GMCS protocol \cite{grosshans2003quantum}. We assume, as usual, that Eve has complete control of the polarization transformation occurring between Alice and Bob. She also knows the polarization parameters $\theta$, $\phi$, and $\delta$ exactly. This capability and knowledge does not give her any additional information about the GMCS raw or secret key. Once all parties know the polarization parameters and Alice transforms her stored encodings digitally, Alice, Bob, and Eve arrive at the standard GMCS scenario. In this case there are two GMCS channels in orthogonal polarizations, but they are completely independent.  %Thus this could be considered a two-mode independently encoded multiplexing approach that sacrifices half of the potential rate to simplify the receiver. 

\section{Discussion}
We have presented a polarization agnostic encoding method for GMCS CVQKD. Our method removes the need to actively manipulate the optical signal to mitigate polarization changes in the optical link between Alice and Bob when monitoring a single polarization. In doing so, we eliminate the cost and loss associated with polarization control components. We also avoid noise injected by polarization correction algorithms. 
%We note that by using our method and \emph{true} heterodyne techniques it may be possible to achieve a passive polarization management approach with a single balanced detector \cite{brunner2017low, pereira2023polarization}. 
Future work will investigate real-world deployment of our method and integration of passive encoding methods \cite{qi2018passive,qi2020experimental}.

\begin{acknowledgments}
This work was performed at Oak Ridge National Laboratory, operated by UT-Battelle for the U.S. Department of Energy under Contract No. DEAC05-00OR22725. Funding was provided by the  U.S. Department of Energy under the Risk Management Tools and Technologies program in the Broadly Applicable Quantum Key Exchange Network (BAQKEN) project.
\end{acknowledgments}

\bibliographystyle{apsrev4-1}
\bibliography{CVQKDpolAgnostic.bib}

\end{document}